\documentclass[10pt,twocolumn]{revtex4}
\usepackage{graphicx}
\usepackage{psfig}
\newcommand{\be}{\begin{equation}}
\newcommand{\bea}{\begin{eqnarray}}
\newcommand{\ee}{\end{equation}}
\newcommand{\eea}{\end{eqnarray}}
\newcommand{\eps}{\epsilon}

\begin{document}

\title{Directed transport in a classical lattice with a high-frequency driving}
\author{A.P. Itin$^{1,2}$ and A.I.Neishtadt$^{2,3}$}
\affiliation{$^{1}$Zentrum f\"{u}r optische Quantentechnologien,
Universit\"{a}t Hamburg, Luruper Chaussee 149, 22761 Hamburg, Germany\\
$^2$Space Research Institute (IKI), Russian Academy of Sciences,
117997, 84/32 Profsoyuznaya Str, Moscow, Russia\\
$^{3}$Department of Mathematical Sciences, Loughborough
University, Loughborough, LE11 3TU, UK. }

\begin{abstract}

We analyze the dynamics of a classical particle in a spatially
periodic potential under the influence of a periodic in time
uniform force. It was shown in [S.Flach, O.Yevtushenko, Y.
Zolotaryuk, Phys. Rev. Lett. {\bf 84}, 2358 (2000)] that despite
zero average force, directed transport is possible in the system.
Asymptotic description of this phenomenon for the case of slow
driving was developed in [X. Leoncini, A. Neishtadt, A. Vasiliev,
Phys. Rev. E {\bf 79}, 026213 (2009)]. Here we consider the case
of fast driving using the canonical perturbation theory. An
asymptotic formula is derived for the average drift velocity as a
function of the system parameters and the driving law. We show
that directed transport arises in an effective Hamiltonian that
does not possess chaotic dynamics, thereby clarifying the relation
between chaos and transport in the system. Sufficient conditions
for transport are derived.
\end{abstract}

\maketitle

Transport phenomena in nonlinear systems have attracted growing
interest in the recent decades \cite{Reiner, Turaev,
Casati,Chaos,Flach,Flach2}. In particular, directed transport in
periodic potentials $U(x+2\pi)=U(x)$ under the influence of an
unbiased external force $E(t)$ has been subject of research in
numerous papers recently. The studies of the transport are
motivated, e.g., by  its prospects for various technological
applications \cite{Reiner}. There is also much related activity in
the quantum realm \cite{Alberti, Holthaus,Monteiro,Kolovsky}.

Let us concentrate on Hamiltonian systems (no dissipation).
Symmetry analysis allows one to formulate {\em necessary}
conditions of existence of directed motion in an ensemble of
particles with zero average initial velocity \cite{Flach,Flach2}.
However, what are {\em sufficient} conditions to be imposed on
$E(t)$ and $U(x)$ to guarantee the transport? What is the average
velocity of transport in an {\em arbitrary} periodic potential
$U(x)$ and force $E(t)$? Despite a lot of research and
publications in this field, it seems that it is not possible to
derive explicit analytical expressions for average velocity of a
particle as a function of system's parameters, apart from the two
cases: slow or fast driving, where the frequency of perturbation
is much smaller or faster than the unperturbed frequency of the
system, respectively. In these cases, it is possible to apply
methods of classical perturbation theory \cite{AKN}. In
\cite{Neish09}, classical adiabatic theory was applied to the case
of slowly time-dependent force. Here we develop classical
perturbation analysis of the opposite, high-frequency limit, which
is more interesting from an experimental point of view. We derive
a general formula for the average velocity of particles in a
periodic potential under a high-frequency drive of an arbitrary
form, which provides us with {\em sufficient} conditions for the
transport.

We consider the system with the Hamiltonian \be H = \frac{P^2}{2}
+ U(x) - x E( \omega t), \ee i.e. a particle in a spatially
periodic potential ($U(x+2\pi)=U(x)$) influenced by a spatially
uniform force $E( \omega t)$ which is periodic in time and has
zero mean: $E(\omega t + 2\pi) = E(\omega t)$. We assume the force
is changing fast: $\eps \equiv \frac{1}{\omega} \ll 1$. Since the
potential is defined up to a constant, we fix $ \langle U(x)
\rangle_x =0$, where $ \langle .. \rangle_{\mu}$ means averaging
over the variable $\mu$. Equations of motion are $ \dot{x}= P,
\quad \dot{P}= - \partial U/\partial x + E(t/\eps). $

We apply canonical perturbation theory, shifting time-dependence
to higher order terms in $\eps$. The main idea is to obtain an
effective time-independent Hamiltonian, and then to determine how
the initial distribution of phase points is located in the phase
space of this new Hamiltonian. Importantly, the effective
Hamiltonian does not possess chaotic dynamics.

We start with several preliminary transformations. Introducing the
fast time $\tilde t = t/\eps$ ( $ \frac{df}{d \tilde t}= \eps
\dot{f}$), the new Hamiltonian is (the tilde over the new time is
omitted from now on) $ \eps H = \eps [ \frac{P^2}{2} + U(x) - x
E(t) ]. $

We make a canonical transformation $P,x \to p,x$ using a
generating function $W_1 (p,x,t) = x(p + \eps f(t))$. The new
Hamiltonian is
$$ \eps \hat H = \eps  H + \eps \frac{\partial W_1}{\partial t} =
\eps [\frac{(p+\eps f)^2}{2} +U(x) - x( E(t)- \frac{df}{d t})]. $$
In the following, let us omit the hat over the Hamiltonian, which
we denote as an 'intermediate'. We choose $f(t)= \{E\},$ where
$\{..\}$ denotes an integrating operator: $ \{ F \} \equiv
\int_{t_0}^t F(s)ds - \langle \int_{t_0}^t F ds \rangle_t$. Then,
the intermediate Hamiltonian is \be \eps H = \eps \Bigl[ \frac{(p
+\eps f(t))^2}{2} + U(x) \Bigr]. \ee

We finally make a canonical transformation $ (x,p) \to ( \bar x,
\bar p)$ using the generating function $$ W_2 =x \bar p + \eps^2
S(x, \bar p, t) \equiv x \bar p + \eps^2 S_1(x, \bar p, t) +
\eps^3 S_2(x, \bar p,t ) + .., $$ where all functions $S_i$ (to be
determined later) are periodic in time. Variables and the
Hamiltonian are transformed as ($\frac{\partial S_i}{\partial p}$
denotes differentiation over new momentum $\bar p$):

\bea \bar x = x + \eps^2 \frac{\partial S_1}{\partial p}+\eps^3
\frac{\partial S_2}{\partial p}&+&..,\quad p=\bar p + \eps^2
\frac{\partial S_1}{\partial x}+\nonumber\\
+\frac{\partial S_2}{\partial x}+.., \qquad {\cal H}(\bar p, \bar
x, \eps) &=&  H(p,x,t,\eps) + \eps \frac{\partial S}{\partial t},
\label{transform}
 \eea
where the new Hamiltonian is $\eps {\cal H}$, and we denote it as
an 'effective'. We now expand the effective ($\eps {\cal H}$) and
the intermediate ($\eps H$) Hamiltonians in powers of $\eps$, and
compare the terms with the same powers of $\eps$. This give us
equations defining the generating function $S$ (see Supplementary
information for details):

\bea {\cal H}(\bar p, \bar x, \eps) &=& {\cal H}_0(\bar p, \bar x
)+ \eps {\cal H}_1(\bar p, \bar x ) +  \eps^2 {\cal H}_2(\bar p,
\bar x )+.. \nonumber\\&=& H(p,x,t,\eps) + \eps \frac{\partial
S}{\partial t} = H_0(p,x,t)
\\+\eps( H_1(p,x,t)&+&\frac{\partial S_1}{\partial t} ) + \eps^2(
H_2(p,x,t) + \frac{\partial S_2}{\partial t} )+.. \nonumber\eea

We obtain ${\cal H}_1={\cal H}_2={\cal H}_3=0$, ${\cal H}_4=
\frac{1}{2}  \langle v_1^2 \rangle_t U''(x) $, where $v_1 \equiv
\{f\}$.

We see that the effective Hamiltonian coincides with the
unperturbed one up to the fourth order of the perturbation theory.
The expressions for all terms of the generating function up to the
fourth order can be found in Supplementary information. These
expressions are useful for studying various phenomena in periodic
potentials with high-frequency driving.

\begin{figure}
\includegraphics[width=34mm]{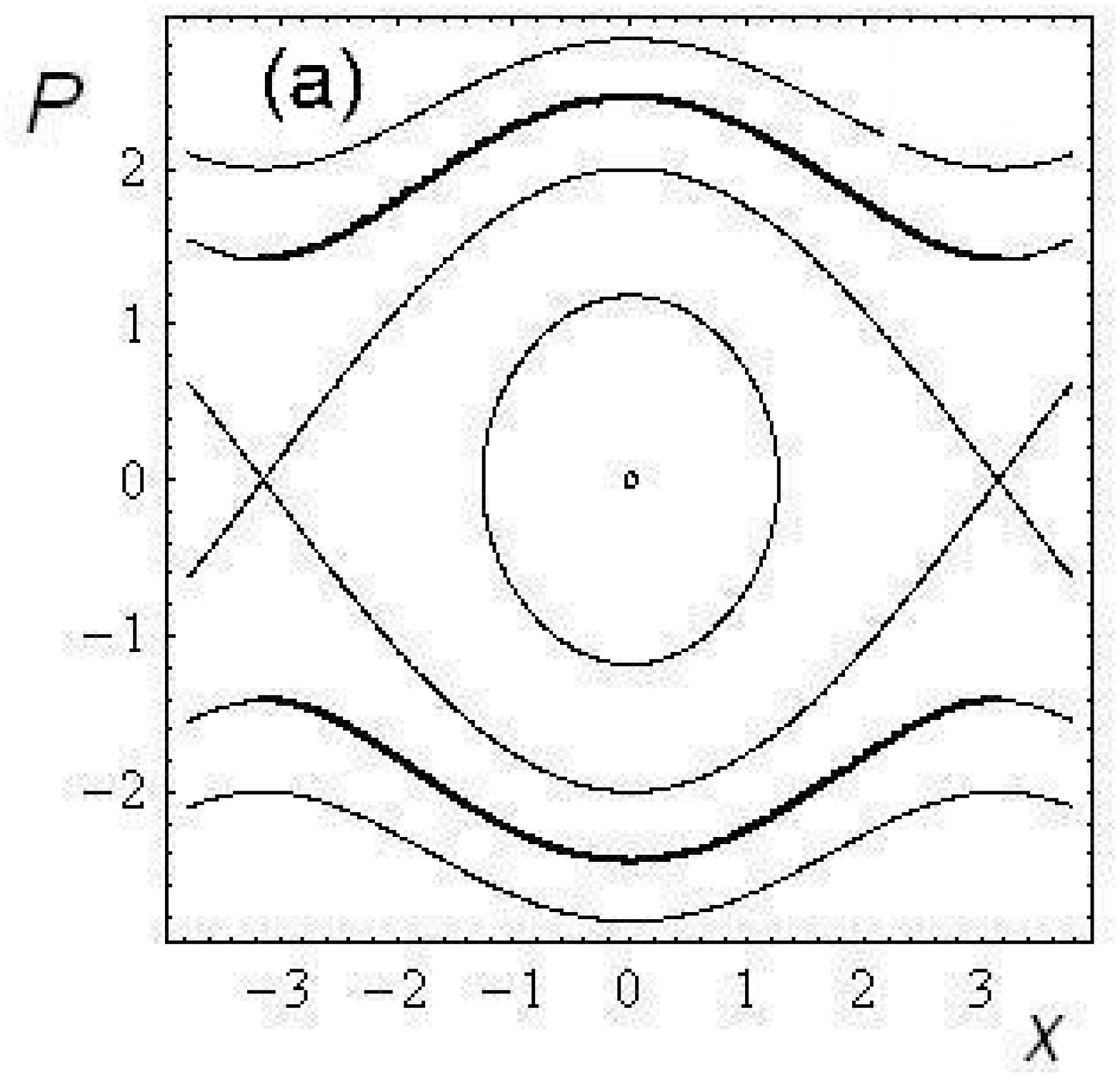}
\includegraphics[width=42mm]{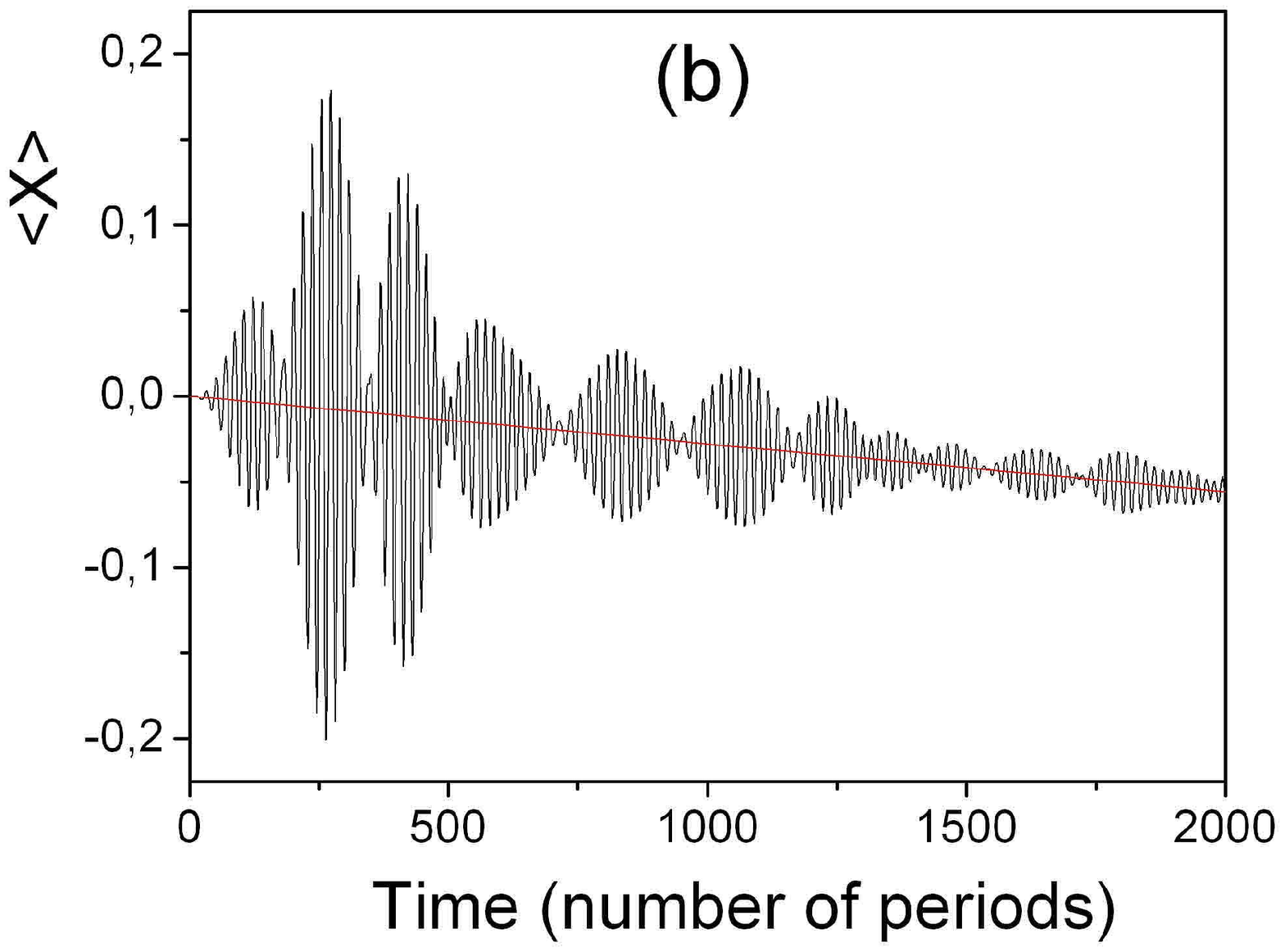}
\caption{ (a) Phase portrait of the unperturbed Hamiltonian and
the initial distribution of particles (two segments of
trajectories drawn in bold thick curves).
 (b) Drift of the average position of ensembles of
particles with time in the potential $U(x)=-\cos x$ under the
bichromatic perturbation described in the main text. We integrate
100 ensembles of particles, each ensemble contains 200 particles
being distributed along two symmetric unperturbed trajectories
with ${\cal{E}}_0=2.0$. Distribution is uniform in canonical
'angle' variable of the unperturbed Hamiltonian. To each ensemble,
we apply a force $E(t-t_i)$, where $t_i$ is distributed uniformly
on $(0,2 \pi)$. Averaging over all ensembles gives us the average
position $\bar x (t)$ (for convenience, a normalized position
$\langle X \rangle= \bar x/2\pi \eps $ is shown, where the
normalization is  such that the slope of $\langle X \rangle(t)$,
with time measured in units of fast periods, gives
 the average velocity). The inverse frequency $\eps=0.03$. After a certain transient
 period with beating oscillations of the average positions, the asymptotic regime
 is established. \label{T1}}
\end{figure}

Now, for studying directed transport, we consider an ensemble of
particles distributed along two unperturbed symmetric trajectories
with the same energy $E \equiv H_0 = {\cal E}_0$ (see Fig.
\ref{T1}a). Distribution of the particles is uniform in canonical
'angle' variable of the unperturbed Hamiltonian (introducing
action-angle variables, constant energy implies constant action
and the uniform distribution over the 'angle' variable seems to be
natural). We abruptly apply (at a certain initial moment $t_0$)
the perturbing force to this distribution. We shall now derive a
formula for average velocity of particles in the ensemble,
averaging over the initial time $t_0$ as well. Procedure for this
is as following. After making the canonical transformations
described above, we obtain the time-independent effective
Hamiltonian where average velocity of each particle is easy to
calculate. The transformations depend on the initial time $t_0$.
Because of the asymmetries in the function $f(t)$, initial
symmetric distribution will not be symmetric in the new variables.
This will give us the transport velocity, nontrivial part of which
will survive even after averaging over $t_0$. Quantitatively, we
will have
\bea \bar v &=& \frac{v_+ + v_-}{2}, \quad v_{\pm} = \label{transport}\\
 &=& \int \limits_0^{2\pi} \int \limits_0^{2\pi}
 \Theta ( E_{\pm}-E_s) \mbox{sgn}[\bar p_{\pm}(x, {\cal{E}}_0)] \frac{\bar \omega (E_{\pm})
 \omega_0 ({\cal{E}}_0)}{(2 \pi)^2 |p(x,{\cal{E}}_0)|} dt dx
 \nonumber
 \eea
Here, $v_{\pm}$ denote contributions from the unperturbed
trajectories $H= {\cal{E}}_0$ with positive and negative 'old'
momentum, correspondingly, $\bar p_{\pm}(x)$ are the new
(transformed) momenta of the particles from these trajectories.
$E_{\pm}$ are the new energies of these phase points
$(E_{\pm}={\cal H})$ in the effective Hamiltonian, and $E_s$ is
the energy of the separatrix of the new Hamiltonian dividing areas
of bounded and unbounded motion; $\Theta(x)$ is the Heaviside step
function (particles coming into the bounded region of phase space
do not contribute to transport). Then, $\bar \omega$ is the
frequency of canonical 'angle' variable in the new Hamiltonian
(which gives contribution to the transport velocity), while the
term $\omega_0/|p(x,{\cal E}_0)|$ gives us distribution of
particles in $x-$coordinate (recall that we consider uniform
distribution in 'angle' variable of the unperturbed Hamiltonian).
Both frequencies we define using 'old' time, so that the drift
velocity also is defined using 'old' time.

For simplicity, we consider initial energies not very close to the
separatrix, ${\cal{E}}_0-{\cal{E}}_s \gg \eps$. Then the integrand
of Eq.(\ref{transport}) simplifies to $\pm \frac{\bar \omega
(E_{\pm})
 \omega_0 ({\cal{E}}_0)}{(2 \pi)^2 |p(x,{\cal{E}}_0)|}$.


Expressions for $\bar p, \bar x$ needed in Eq.(\ref{transport})
are given by Eq.(\ref{transform}), with terms up to 4th order
being presented in Supplementary information. Expanding
Eq.(\ref{transport}) in powers of $\eps$ and using the
above-mentioned expressions, we get after double integration over
$x$ and the initial time,  in the lowest order in $\eps$, \be \bar
v = - \eps^3 \langle f^3 \rangle_t \omega_0({\cal E}_0) \Bigl[
\frac{1}{2}\frac{\partial^2 \omega_0}{\partial {\cal E}^2}+
\frac{1}{3} \frac{ \partial^3 \omega_0  }{\partial {\cal
E}^3}{\cal E}_0 \Bigr]. \label{formula} \ee This is one of the
main results of this Letter. For the particular case of $U(x)=-
\cos x$ we obtain, explicitly, \bea  \bar v &=& \frac{\eps^3
\langle f^3 \rangle_t \pi^2}{48(1+ {\cal{E}}_0)({\cal{E}}_0-1)^3
\mbox{K}^5(\kappa) } \Bigl[
-6(1+{\cal{E}}_0)^2 {\cal{E}}_0 \mbox{E}^3(\kappa) \nonumber\\
&+& 6(1+{\cal{E}}_0)(1+3{\cal{E}}_0^2)
\mbox{E}^2(\kappa)\mbox{K}(\kappa) \label{asymptotic}\\ &-&
{\cal{E}}_0(15+17{\cal{E}}_0^2)\mbox{E}(\kappa)\mbox{K}^2(\kappa)
 +({\cal{E}}_0-1)(3+5{\cal{E}}_0^2)\mbox{K}^3(\kappa)
 \Bigr], \nonumber \eea where $\mbox{K,E}$ are elliptic
integrals of the first and second type, correspondingly, and  $
\kappa = \frac{2}{1+{\cal{E}}_0}$. The expression is valid for all
initial energies not very close to the separatrix (not necessarily
large ones). Let us compare this result with the earlier results
of \cite{Flach} (obtained for the same potential $U(x) = -\cos x$,
but for large energies). In the limit of large energies, we have

\be  \bar v \approx -\frac{15 \eps^3 \langle f^3 \rangle_t }{16
{\cal{E}}_0^3}. \ee

For the perturbation used in \cite{Flach}, $E = E_1 \cos(\omega t
) + E_2 \cos(2 \omega t + \alpha)$, one has $\langle f^3 \rangle =
-\frac{3}{8} E_1^2 E_2 \sin \alpha $ and we have \be \bar v =
\frac{45}{16} \frac{\eps^3 E_1^2 E_2 \sin \alpha}{P_0^6},
 \ee
which reproduces the result of \cite{Flach} up to the constant
factor, which is different (note that, in the limit of large
energies, the distribution which is uniform in canonical 'angle'
becomes uniform in coordinate $x$).

Note that in the 'new' (fast) time, the drift velocity is of the
order of $\eps^4$. That is why we choose to use the fourth-order
perturbation theory, even though in the final expressions
(\ref{formula},\ref{asymptotic}) even $v_1$ is not needed. In the
leading term, the force is entered only via $\langle f^3
\rangle_t$, which provides a {\em sufficient} condition for the
nonzero drift velocity of the third order of $\eps$: $\langle f^3
\rangle_t \ne 0$ (at the energy level where $\Bigl[
\frac{1}{2}\frac{\partial^2 \omega_0}{\partial {\cal E}^2}+
\frac{1}{3} \frac{ \partial^3 \omega_0  }{\partial {\cal
E}^3}{\cal E}_0 \Bigr]=0$ transport will be strongly supressed,
but this condition only influence certain specific trajectories).
Meanwhile, the arguments based on symmetries of the perturbation
can only give {\em necessary} conditions for non-vanishing
transport.


\begin{figure}
\includegraphics[width=42mm]{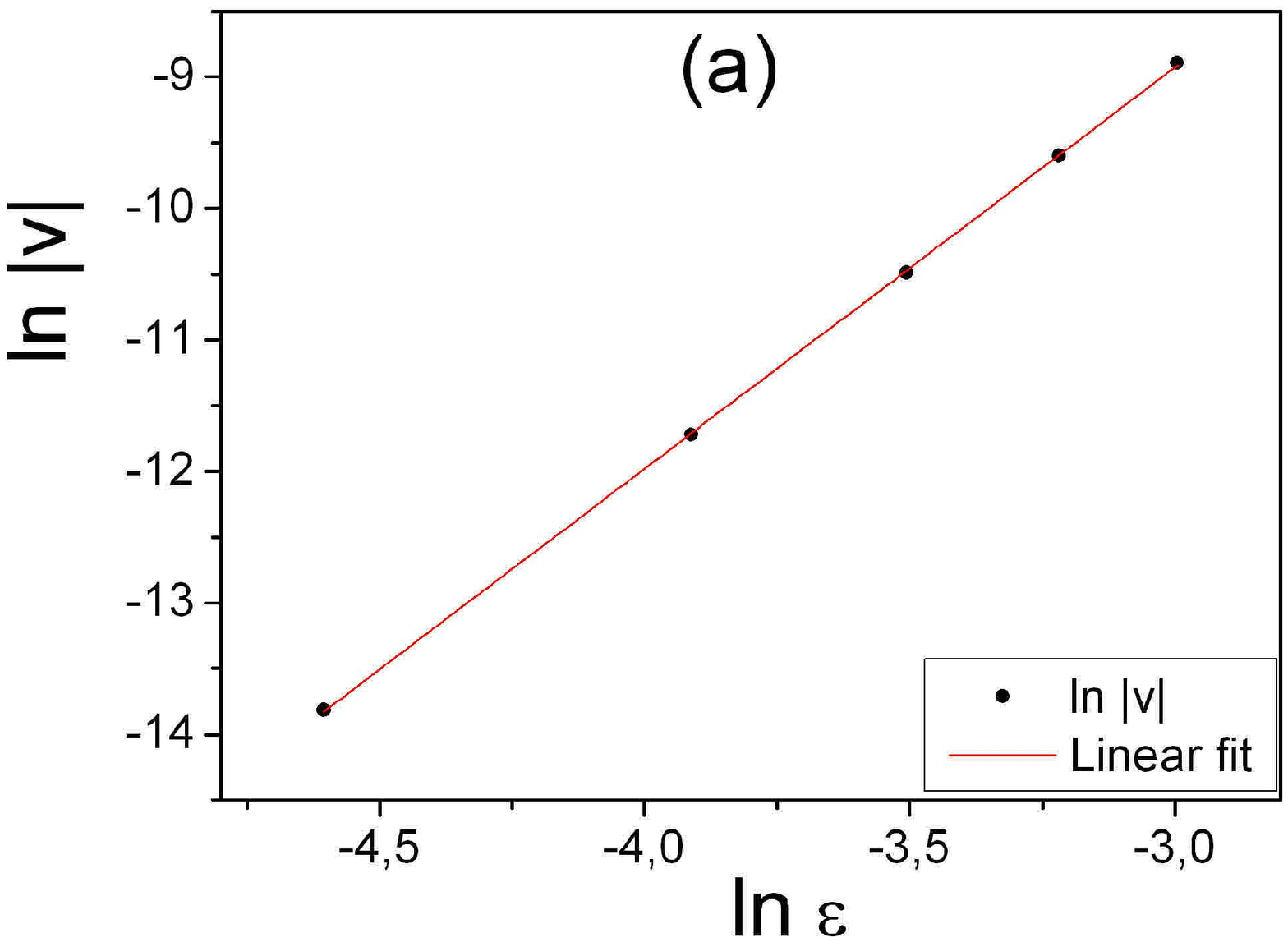}
\includegraphics[width=42mm]{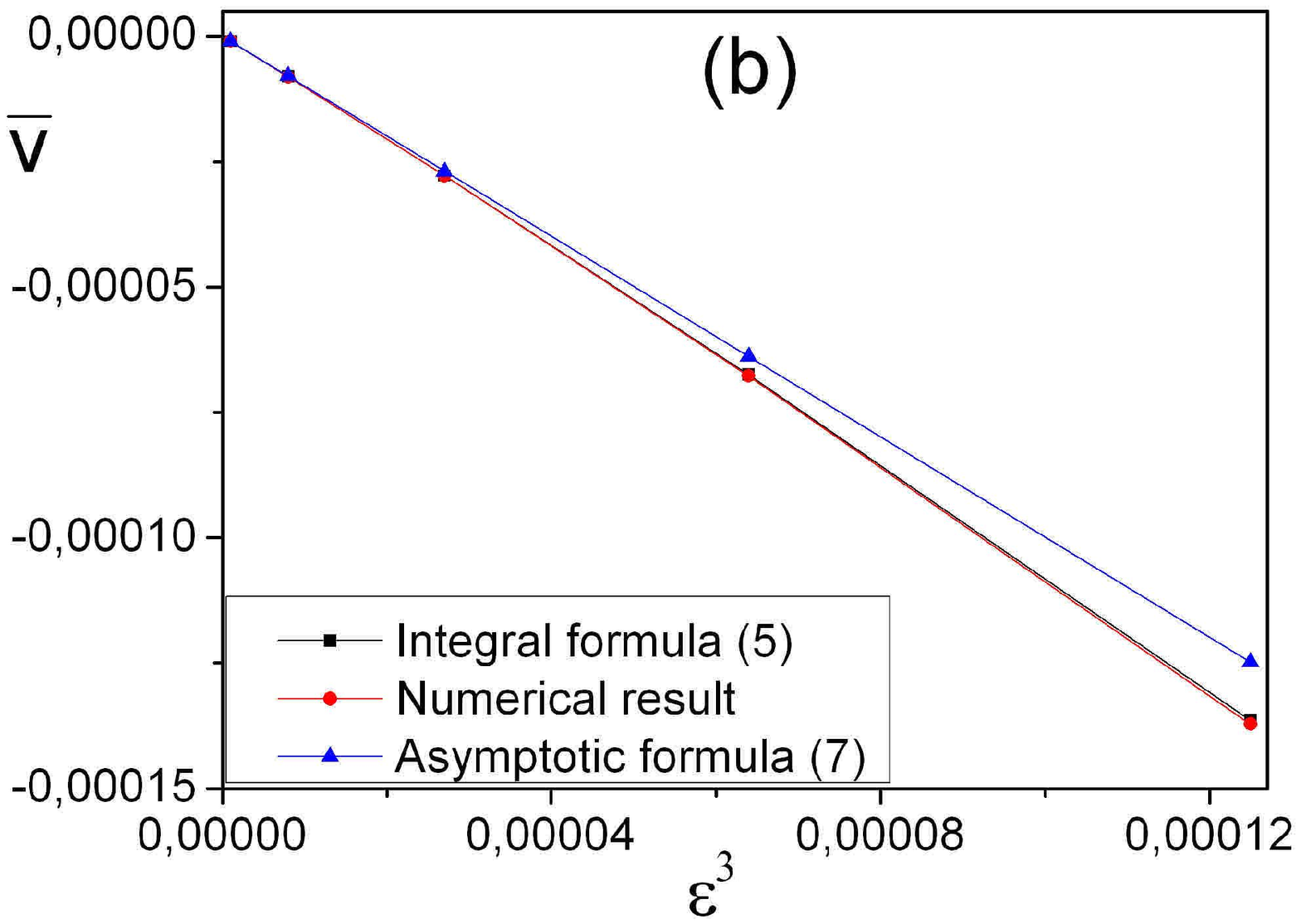}
\caption{(a)Dependence of the average drift velocity on the
inverse frequency $\eps$ for $U(x)=-\cos x$ and the perturbation
described in the main text. Linear fitting give the coefficient
$\alpha=3.05$ for the power-law $\ln |v|= \alpha \ln \eps$.
Deviation from the asymptotic value $\alpha=3.0$ is partly due to
the higher-order terms which are not so small in this range of
$\eps$. (b)  Average velocity as a function of $\eps^3$. Squares:
integral formula of Eq. (\ref{transport}). Circles: numerical
results. Triangles: asymptotic formula of Eq. (\ref{asymptotic}).
Lines are guides to the eye. Numerical results and the integral
formula are almost indiscernible at this scale. The asymptotic
formula deviates from numerics at larger values of $\eps$, where
higher order terms become significant.
 \label{T2}}
\end{figure}



Numerically, we prepare a symmetric distribution of initial
conditions in the phase space of the unperturbed Hamiltonian, and
(abruptly) apply the force with an offset $t_i$. Applying the
time-dependent force $E(t-t_i)$, it is important to average over
the initial phase $t_i$. Otherwise, one can obtain directed
transport even in the case of harmonic driving \cite{Flach2}, as a
function of the initial phase. In detail, our procedure is as
follows. We prepare $N$ copies of a symmetric in coordinate and
momentum initial phase-space distribution. Specifically, we choose
as an initial distribution a collection of phase points
distributed over two symmetric trajectories of the unperturbed
Hamiltonian with the energy ${\cal E}= {\cal{E}}_0$. To each copy,
we apply at $t=0$ a force with an initial offset $t_i$: $E_i(t)=
E(t-t_i)$, where $E_i(t)$ is the force acting on the $i-$th copy
of the phase space, and $t_i$ is an offset uniformly distributed
on $(0,2\pi)$. We average over all phase points in all $N$ copies
of the phase space.  In other words, we average not only over
initial phase-space distribution, but also over initial phase of
the force.

For a numerical example, we consider the potential $U(x)=-\cos x$
and the following perturbation: $ E(t) = -V_1 \sin t - 2V_2
\sin(2t+\pi/4), $ where $V_1 = 3, \quad V_2 = 0.8$. This
corresponds to \be f(t) = V_1 \cos t + V_2 \cos(2t + \pi/4), \ee
and we have: $ v_1(t) =  V_1 \sin t + \frac{V_2}{2}
\sin(2t+\frac{\pi}{4}) ,\quad \langle v_1^2 \rangle_t =
\frac{V_1^2}{2}+\frac{V_2^2}{8},$  $\langle f^3 \rangle_t =
\frac{3V_1^2 V_2}{4\sqrt{2}}. $ We distribute particles along the
trajectory of the unperturbed Hamiltonian with ${\cal E}_0=2$.

In Table I, we compare the numerical results for the average
velocity with predictions of the integral formula
(\ref{transport}) and the asymptotic formula (\ref{asymptotic}).
In Fig.\ref{T1}b, time evolution of the mean coordinate of
ensembles of particles is shown for $\eps=0.03$. Slope of the
fitted line gives the average velocity. In Fig.\ref{T2}a, the
dependency of the average drift velocity on $\eps$ is shown. Note
that the velocity is defined using 'old', original time. In Fig.
\ref{T2}b, predictions of the integral formula (\ref{transport}),
the asymptotic formula (\ref{asymptotic}) and numerical
calculations are compared.

\begin{table}[ht]
\caption{Average velocity from numerics and theory}
\begin{tabular}{| c | c | c | c |}
\hline \hline $\eps $  & Numerical result  & Integral Eq.
(\ref{transport}) & Asymptotic Eq. (\ref{asymptotic})  \\
\hline
0.01 & -1.004 $ \cdot 10^{-6}$ & -1.002 $ \cdot 10^{-6}$ & -9.987 $ \cdot 10^{-7}$ \\
0.02 & -8.130 $ \cdot 10^{-6}$ & -8.099 $ \cdot 10^{-6}$ & -7.990 $ \cdot 10^{-6}$ \\
0.03 & -2.784 $ \cdot 10^{-5}$ & -2.781 $ \cdot 10^{-5}$ & -2.697 $ \cdot 10^{-5}$ \\
0.04 & -6.782 $ \cdot 10^{-5}$ & -6.756 $ \cdot 10^{-5}$ & -6.392 $ \cdot 10^{-5}$ \\
0.05 & -1.372 $ \cdot 10^{-4}$ & -1.364 $ \cdot 10^{-4}$ & -1.248 $ \cdot 10^{-4}$ \\
 \hline
\end{tabular}
\end{table}

To conclude, we derive the formula for average velocity of
particles in a periodic potential under the influence of an
unbiased high-frequency force. The average velocity is related to
a certain integral momentum of the force, thereby the formula is
useful for a wide range of perturbations including (but not
limiting to) bichromatic or multichromatic harmonic driving,
anharmonic driving, etc. Moreover, it provides {\em sufficient}
conditions for directed transport, thereby conditions based on
symmetry of the driving only give {\em necessary} conditions of
existence of directed transport. Conceptual aspects of the
phenomenon are clarified now. Indeed, applying classical
perturbation theory of the 4-th order, one obtains an effective
time-independent Hamiltonian. Although the Hamiltonian is
symmetric in momentum, drift of particles occurs because the
canonical transformation leading to the new Hamiltonian
asymmetrically transforms original particle distribution into the
distribution of particles in the new variables. A priori, it was
not clear that mechanism of transport should be like this. In
higher orders of the perturbation theory, one obtains an effective
Hamiltonian with asymmetry in momentum \cite{Semenova}. It might
be possible that this asymmetry would be responsible for the
transport, which is shown not to be the case: generally, transport
arises within the 4th order of the perturbation theory.
Importantly, the effective Hamiltonian does not possess chaotic
dynamics. The full Hamiltonian has only a narrow stochastic layer
on its phase space in the vicinity of the separatrix of
unperturbed Hamiltonian. This layer becomes exponentially small in
the high-frequency limit and cannot be described in any finite
order of the perturbation theory. Since we consider energies not
very close to the separatrix, this chaotic layer is completely
irrelevant for the dynamics. In other words, chaos is not needed
for the directed transport.


Procedure of the averaging used in our work is very important in
the high-frequency case. Without averaging over the initial phase,
it is not possible to catch the non-trivial part of the transport.
That is, considering perturbation with a fixed initial phase and
averaging only over the phase space, one gets directed transport
even with the simple perturbation $E(t)=\cos (\omega t-\phi_0)$.
So far, experiments \cite{Renzoni} only probed transport at fixed
initial phase of the perturbation \cite{RenzoniPlus}. We here
reveal essential features of the nonlinear transport in the
high-frequency regime that has not been probed in the experiments
yet, and develop analytical theory for them. The theory can be
probed in the experiments similar to \cite{Renzoni}, but with a
high-frequency perturbation. Our work can therefore inspire new
experiments in this field.

This work was supported in part by RFBR 09-01-00333 and
NSh-2519.2012.1 grants. The authors are grateful to L.Semenova for
participation in the research on early stages, and to C. Petri,
I.Brouzos, C.Morfonios, S.Flach, P.Schmelcher for useful
discussions.

\newpage
\section*{Supplementary information}

We expand the effective Hamiltonian in series in $\eps$:

\bea {\cal H}(\bar p, \bar x, \eps) &=& {\cal H}_0(\bar p, \bar x
)+ \eps {\cal H}_1(\bar p, \bar x ) +.. \\  &=& {\cal H}(\bar p, x
+\eps^2
\frac{\partial S}{\partial p}, \eps)  \nonumber\\
 &=& {\cal H}_0 (\bar p,  x) + \frac{\partial {\cal H}_0 }{ \partial x } \Bigl( \eps^2 \frac{\partial S_1}{\partial p} +
  \eps^3 \frac{\partial S_2}{\partial p} +..\Bigr) + \nonumber\\
  &+& \frac{1}{2} \frac{\partial^2 {\cal H}_0 }{\partial x^2} \Bigl( \eps^2 \frac{\partial S_1}{\partial p} +
  \eps^3 \frac{\partial S_2}{\partial p} +.. \Bigr)^2 + .. \nonumber\\
  &+&\eps {\cal H}_1 (\bar p, x )
  + \eps \frac{\partial {\cal H}_1 }{ \partial x } \Bigl( \eps^2 \frac{\partial S_1}{\partial p} +
  \eps^3 \frac{\partial S_2}{\partial p} +..\Bigr) + \nonumber\\
 &+& \frac{1}{2} \eps \frac{\partial^2 {\cal H}_1 }{\partial x^2} \Bigl(
 \eps^2 \frac{\partial S_1}{\partial p} +
  \eps^3 \frac{\partial S_2}{\partial p} + .. \Bigr)^2 + ..\nonumber\\
   &+& \eps^2  {\cal H}_2 (\bar p,  x) +
 \eps^2 \frac{\partial {\cal H}_2 }{ \partial x } \Bigl( \eps^2
\frac{\partial S_1}{\partial p} +
  \eps^3 \frac{\partial S_2}{\partial p} +..\Bigr).. \nonumber
     \eea

On the other hand, the effective and the intermediate Hamiltonians
are related as \bea {\cal{H}} (\bar p, \bar x, \eps) &=&
H(p,x,t,\eps) + \eps \frac{\partial S}{\partial t} \\
&=& H_0(p,x,t) + \eps( H_1(p,x,t) + \frac{\partial S_1}{\partial
t} ) \nonumber\\
&+& \eps^2( H_2(p,x,t) + \frac{\partial S_2}{\partial t} )+..
\nonumber\\ &=& H_0(\bar p,x,t) + \frac{\partial H_0}{\partial p}
\Bigl( \eps^2 \frac{\partial S_1}{\partial x}  + \eps^3
\frac{\partial S_2}{\partial x}+... \Bigr) \nonumber\\ &+&
\frac{1}{2}\frac{\partial H_0^2}{\partial p^2} \Bigl( \eps^2
\frac{\partial S_1}{\partial x}+ \eps^3 \frac{\partial
S_2}{\partial x} + .. \Bigr)^2  \nonumber\\ &+& \eps( H_1(\bar
p,x,t) + \frac{\partial S_1}{\partial t} ) + \eps \frac{\partial
H_1}{\partial p}\Bigl( \eps^2 \frac{\partial S_1}{\partial x}+
..\Bigr) \nonumber\\ &+& ... \nonumber\eea

We have

$H= \frac{p^2}{2} +U(x) + \eps p f(t) + \eps^2 \frac{f(t)^2}{2},
\quad$ $H_0 = \frac{p^2}{2} + U(x), \quad H_1= p f(t), \quad H_2 =
\frac{f(t)^2}{2}.$

The term $\frac{f(t)^2}{2} $ does not influence the dynamics and
can be safely omitted.

Comparing terms of the same order in $\eps$, we have (bars over
$p$ are omitted):
\newline
The zero-order terms in $\eps$: \be \eps^0: \quad {\cal H}_0(p,x)=
H_0(p,x)=\frac{p^2}{2}+ U(x),\nonumber \ee
\newline The first-order terms in $\eps$:
\bea \eps^1: \quad {\cal H}_1 &=& H_1+ \frac{\partial S_1
}{\partial t}, \quad  {\cal H}_1 = p \langle f \rangle_t = 0, \\
S_1 &=& -p v_1(t), \quad v_1(t) = \{f \}. \nonumber \eea
\newline The second-order terms:

\bea \eps^2: \quad {\cal H}_2 &+& \frac{\partial H_0}{\partial x}
\frac{\partial S_1}{\partial p} = H_2 + \frac{\partial S_2
}{\partial t} + \frac{\partial H_0}{\partial p} \frac{\partial S_1
}{\partial x}, \nonumber\\
\quad {\cal H}_2  &=&  0,
 \quad S_2 = -v_2(t) U'(x), \nonumber\\
v_2(t) &=& \{v_1\}  \eea
\newline The third-order terms: \bea \eps^3: \quad {\cal H}_3 =0, \quad S_3
= p v_3(t) U''(x), \quad v_3(t)= \{ v_2 \} \eea
\newline The fourth-order terms :
 \bea \eps^4: \quad  {\cal H}_4 &+& \frac{\partial {\cal H}_0}{\partial
 x}\frac{\partial S_3}{\partial p} + \frac{1}{2}\frac{\partial^2 {\cal H}_0}{\partial
 x^2}\Bigl( \frac{\partial S_1}{\partial p} \Bigr)^2 \nonumber\\ &=& \frac{\partial S_4}{\partial
 t} + \frac{\partial H_0}{\partial p} \frac{\partial S_3}{\partial
 x}+  \frac{\partial H_1}{\partial p} \frac{\partial S_2}{\partial
 x}+ \frac{1}{2} \frac{\partial H_0^2}{\partial p^2} \Bigl(
 \frac{\partial S_1}{\partial x}
 \Bigr)^2, \nonumber\\
  \quad {\cal H}_4 &=& (- \langle f
v_2 \rangle_t -\frac{1}{2} \langle v_1^2 \rangle_t)U''(x) =
\frac{1}{2} \langle v_1^2 \rangle_t U''(x). \nonumber \eea


\begin{thebibliography}{10}
\bibitem{RenzoniPlus} In \cite{Renzoni} a bichromatic perturbation of the form
$\alpha(t)=\alpha_0[A \cos \omega t + B \cos (2\omega t-\phi)]$
was considered ($\alpha(t)$ is a phase modulation of an optical
lattice in the experiment which is equivalent to the uniform force
in the system we discuss). Experiments were done at moderate
frequencies, comparable to the frequency of unperturbed system. In
this regime chaos is important and our analysis does not apply. By
increasing the frequency of perturbation, one enters the regular
regime which we describe, where there is no mixing and it is
necessary to average over the initial phase to get the nontrivial
part of the transport (i.e., to consider $\alpha(t)=\alpha_0[A
\cos \omega (t-t_0) + B \cos (2\omega (t-t_0)-\phi)]$ with
different $t_0$ ).
\end{thebibliography}
\end{document}